\documentclass[ aps, pra, twocolumn, 10pt,showpacs,nofootinbib,superscriptaddress,longbibliography]{revtex4-2}
\usepackage{graphicx,dcolumn,bm,amsmath,bbold,mathtools,bigstrut,nicematrix,tikz,amssymb,mathrsfs,natbib,notoccite,comment,,dsfont,pstricks,xcolor,todonotes}
\usetikzlibrary{fit}
\usetikzlibrary{matrix,decorations.pathreplacing}
\usetikzlibrary{arrows.meta, positioning}

\usepackage{listings}
\definecolor{lightestgray}{rgb}{0.90,0.90,0.90}
\usepackage[unicode=true,
bookmarks=true,bookmarksopen=false,
breaklinks=false,pdfborder={0 0 0},colorlinks=true]
{hyperref}
\usepackage{xcolor}
\definecolor{cblue}{rgb}{0.16, 0.32, 0.75}
\definecolor{cred}{rgb}{0.7, 0.11, 0.11}
\hypersetup{%
	,linkcolor=cred
	,citecolor=cblue
	,urlcolor=black
}

\allowdisplaybreaks

\begin{document}
\title{Error bounds for the Floquet--Magnus expansion and their application to the semiclassical quantum Rabi model}
\author{Anirban Dey}
\affiliation{School of Mathematical and Physical Sciences, Macquarie University,  NSW 2109, Australia}
\affiliation{ARC Centre of Excellence for Engineered Quantum Systems, Macquarie University,  NSW 2109, Australia}
\author{Davide Lonigro}
\affiliation{Department Physik, Friedrich-Alexander-Universit\"at Erlangen-N\"urnberg, Staudtstraße 7, 91058 Erlangen, Germany} 
\author{Kazuya Yuasa}
\affiliation{Department of Physics, Waseda University, Tokyo 169-8555, Japan}
\author{Daniel Burgarth}
\affiliation{Department Physik, Friedrich-Alexander-Universit\"at Erlangen-N\"urnberg, Staudtstraße 7, 91058 Erlangen, Germany} 
\date{\today}
\begin{abstract}
We present a general, nonperturbative method for deriving effective Hamiltonians of arbitrary order for periodically driven systems, based on an iterated integration by parts technique. The resulting family of effective Hamiltonians reproduces the well-known Floquet–Magnus expansion, now enhanced with explicit error bounds that quantify the distance between the exact and approximate dynamics at each order, even in cases where the Floquet–Magnus series fails to converge. We apply the method to the semiclassical Rabi model and provide explicit error bounds for both the Bloch–Siegert Hamiltonian and its third-order refinement. Our analysis shows that, while the rotating-wave approximation more accurately captures the true dynamics than the Bloch–Siegert Hamiltonian in most regimes, the third-order approximation ultimately outperforms both.
\end{abstract}

\maketitle

\section{Introduction}
Quantum systems with time-dependent driving are essential for manipulation, control, and implementation of quantum technologies~\cite{BlochMag, ShirleySol, Mehring, GRIFONI,  Lueffects, Yanbloch, Lauchtbreaking,Gameltime}.  However, solving their exact dynamics is often challenging, even for simple systems, due to the presence of time-dependent terms. Effective Hamiltonians are commonly used to approximate the dynamics~\cite{Gameltime, ZEUCH, Onebound}, yet deriving error bounds between the exact and approximate dynamics is nontrivial and rarely addressed. Providing such bounds is however critical to assess the validity and the accuracy of the approximations~\cite{Onebound, DBtaming, richter2024quantifying,hahn2024efficiency}.

The model that triggered our work is the well-known semiclassical Rabi model, which describes a periodically driven a two-level system (qubit)~\cite{Rabion, Rabispace, eberly, cohenphoton, Wangrotate} with the following Hamiltonian:
\begin{align}
H(t) &= \frac{\omega_{0}}{2}Z + g\cos(\omega t )X\nonumber\\ 
&=  \frac{\omega_{0}}{2}Z + \frac{g}{2}(e^{i\omega t} + e^{-i\omega t})(\sigma^{+}+\sigma^{-}),
\label{rabiham}
\end{align}
where $\omega_{0}\,(>0)$ is the natural frequency of the qubit corresponding to the energy splitting between its two levels, $\omega\,(>0)$ is the driving frequency, and $g$ characterizes the strength of the driving. We assume $g>0$ without loss of generality. Here, $X$, $Y$, and $Z$ are the Pauli matrices, and $\sigma^{\pm} = \frac{1}{2}(X\pm iY)$ are the raising and lowering operators of the two-level system. We set $\hbar = 1$ throughout this work. The Rabi model has been widely studied for its applications in coherent control and quantum state manipulation, making it a cornerstone in the studies of light-matter interactions.

Despite its simplicity, computing its exact dynamics is challenging and the solution is cumbersome~\cite{xie2010,Liang2024,liang2025}. The rotating-wave approximation (RWA)~\cite{Rabion,Rabispace,ScullyZubairy,Lauchtbreaking}, which consists in neglecting the counter-rotating terms $\frac{g}{2}(e^{i\omega t}\sigma^{+} + e^{-i\omega t}\sigma^{-})$, is commonly used to simplify the time-dependence, but breaks down over time even for moderate interaction strengths. Recently, error bounds for the RWA were developed for both the semiclassical~\cite{Onebound} and quantum Rabi models~\cite{DBtaming,richter2024quantifying}, which quantitatively capture its breakdown.

The Bloch--Siegert shift~\cite{eberly, BlochMag, ShirleySol, Giscarddynamics} is a correction to the Hamiltonian in the RWA incorporating the effect of the counter-rotating terms as a perturbation in the inverse frequency. It has been experimentally demonstrated~\cite{pietobs, Zhangbloch,Saikobloch, Liexcitonic, slobosemi}, and it is relevant for achieving precise quantum control and manipulation, particularly in systems such as trapped ions~\cite{Lizuain2008, Warring2013, vedaie2023} and NV centers~\cite{Zhangbloch,Yudilevich_2023,Saikobloch}. The Bloch--Siegert shift can be derived using a variety of methods such as Floquet theory~\cite{Series, ShirleySol, Viebahn}, Magnus--Taylor~\cite{ZEUCH} and Floquet--Magnus expansions~\cite{Casasfloquet,rahaveff,goldmanperiod,goldmandriven,Bukovuniversal,Eckardthigh,MANANGA}, the path-sum approach~\cite{Giscarddynamics}, and by using unitary transformations~\cite{Yanbloch, Lueffects}. While these approaches all predict the correct terms, they do not offer a bound on the error between the dynamics generated by the Bloch--Siegert Hamiltonian and the original Rabi Hamiltonian, which is crucial for determining the validity of the approximation across the range of parameters.

Motivated by the semiclassical Rabi model, we present a nonperturbative general method to derive effective Hamiltonians up to any order for arbitrary bounded periodic Hamiltonians. The key idea is to repeatedly apply an integration-by-part technique originally introduced in Refs.~\cite{Onebound, DBtaming}. A similar idea was recently introduced to bound higher-order versions of the Trotter product formula~\cite{burgarth2024}, and to compute performance bounds in randomized dynamical decoupling~\cite{YiKimMarvian2024}. 
Surprisingly, the family of effective Hamiltonians obtained this way coincides with the one given by the Floquet--Magnus expansion (see, e.g.~Ref.~\cite{Eckardthigh}); additionally, however, our method also provides explicit error bounds at any order, which then apply equally to the Floquet--Magnus expansion. Compared to the tail bound of the Magnus expansion~\cite{BLANESmagnus}, our bounds have two key advantages. Firstly, they do not rely on the convergence of the Magnus expansion and therefore have a wider range of applicability. Secondly, the current bounds on the Magnus expansion come from tail bounds of the exponential function and therefore scale exponentially bad in time, whereas our bounds show linear scaling. Furthermore, while the Floquet–Magnus expansion is limited by a finite radius of convergence, our approach, based on successive refinements of Duhamel’s formula, yields effective Hamiltonians with explicit error bounds without requiring convergence at all. 

By applying this technique to the semiclassical Rabi model, we derive the Bloch–Siegert Hamiltonian and establish error bounds for the approximate dynamics. We compare the dynamics generated by the RWA and Bloch–Siegert Hamiltonians with the original Rabi model, and analyze how their error bounds scale over time. Extending the analysis beyond the Bloch–Siegert approximation, we derive the next-order correction and obtain its explicit error bound. As will be shown, the Bloch--Siegert Hamiltonian is good at capturing the dynamics at multiples of the period, but performs worse than the RWA over time. The third-order correction outperforms both approximations.

The paper is structured as follows. In Sec.~\ref{general iteration}, we introduce our method based on iterated integration by parts for arbitrary time-periodic bounded Hamiltonians, and show the equivalence with the Floquet--Magnus approach. In Sec.~\ref{BS}, we then use this machinery to derive the Bloch--Siegert Hamiltonian and obtain an explicit error bound. In Sec.~\ref{third order}, we derive the third-order effective Hamiltonian and compute its error bound, followed by a comparison with the RWA and Bloch--Siegert Hamiltonians. Finally, we summarize our findings in Sec.~\ref{conclusion}\@. Appendix~\ref{app:bound} contains the proof of one inequality used in Sec.~\ref{general iteration}, while Appendix~\ref{mathematica} includes a Mathematica script for generating effective Hamiltonians of arbitrary orders and computing their error bounds.

\section{General Iterative Integration-by-Part Formula}
\label{general iteration}
In this section we introduce a general method based on an integration-by-part lemma to characterize the difference between two unitary evolutions generated by periodic and bounded Hamiltonians. This method enables us to derive effective Hamiltonians, and to provide rigorous error bounds between the exact and approximate dynamics at any order. Our method is general and non-perturbative: it applies to arbitrary time-dependent Hamiltonians, requires no  assumptions on parameter values, and yields explicit error bounds without requiring convergence. It applies for arbitrary coupling strengths and over any time interval. Its versatility makes it a powerful tool for assessing the validity and the accuracy of effective descriptions.

\subsection{Integral Formula for the Difference between Evolutions}
Our goal is to quantify the distance between the unitary evolutions generated by two distinct time-dependent and bounded Hamiltonians $H_{1}(t)=H_{1}(t+T)$ and $H_{2}(t)=H_{2}(t+T)$ on a Hilbert space $\mathcal{H}$, with common period $T$. We start with the corresponding Schr\"odinger equations, 
\begin{equation}
\frac{d}{dt}U_j(t)=-iH_j(t)U_j(t),\quad U_j(0)=\openone\quad(j=1,2).
\end{equation}
Assuming both $H_1(t)$ and $H_2(t)$ to be locally integrable in time, the difference between the evolutions generated by them can be expressed by a Duhamel integral by directly integrating the corresponding Schr\"odinger equations~\cite{Onebound}:
\begin{multline}
U_{1}(t)-U_{2}(t)\\
= -i\int_{0}^{t} ds\,U_{1}(t) U_{1}^{\dagger}(s)[H_1(s)-H_2(s)]U_{2}(s).
\label{diff01}
\end{multline}
The basic idea is to split the integrand into its time-average and the deviation from the average. To this end, we define the time-average of a time-dependent operator $A(s)$ with period $T$ as
\begin{equation}
\overline{A} =\frac{1}{T}\int_{0}^{T}ds\,A(s),
\end{equation}
and the deviation from average as
\begin{equation}
 \Delta(A(t)) = A(t) - \overline{A}.
\end{equation}
With this notation, we can now write the difference~\eqref{diff01} as
\begingroup
\allowdisplaybreaks
\begin{align}
&
U_{1}(t)-U_{2}(t)
\nonumber\\
&\quad
= -i\int_{0}^{t}ds\,U_{1}(t) U_{1}^{\dagger}(s)\,\Bigl(\overline{H_1-H_2}\Bigr)\,U_{2}(s)\nonumber\\ 
&\quad\quad{}
-i\int_{0}^{t} ds\,U_{1}(t) U_{1}^{\dagger}(s)\Delta(H_1(s)-H_2(s))U_{2}(s),
\label{diff2}
\end{align}
\endgroup
where we drop the time-dependence of the quantity averaged over time.
In the next step, we perform an integration by part replacing the middle term $\Delta(H_1(s)-H_2(s))$ with its integral (see Ref.~\cite{Onebound} for a generalization to locally integrable Hamiltonians) to get
\begingroup
\allowdisplaybreaks
\begin{align}
&
U_{1}(t)-U_{2}(t)\nonumber\\
&
\,
=-iS_t^{(1)}U_{2}(t)-i\int_{0}^{t}ds\,U_{1}(t)U_{1}^{\dagger}(s)\,\Bigl(\overline{H_1-H_2}\Bigr)\,U_{2}(s)\nonumber\\
&
\,\quad{}
-\int_{0}^{t}ds\,U_{1}(t)U_{1}^{\dagger}(s)[H_1(s)S_s^{(1)}-S_s^{(1)}H_2(s)]U_{2}(s),
\label{evodiff}
\end{align}
\endgroup
where we define the integral action by
\begin{equation}
S_t^{(1)} = \int_{0}^{t}ds\,\Delta(H_1(s)-H_2(s)).
\end{equation}
This equation was the key ingredient in Ref.~\cite{Onebound} for deriving a quantitative error bound for the RWA: indeed, the integral action $S_t^{(1)}$ is small for Hamiltonians which oscillate rapidly around the same average and the two evolutions are close to each other. Here we wish to go further and iterate the procedure. 

To this end, we define the time-dependent superoperator $\mathcal{K}_t$ by
\begin{equation}
\mathcal{K}_{t}(A(t)) = H_1(t)A(t)-A(t)H_2(t).
\label{defone}
\end{equation}
We note that this superoperator maps $T$-periodic operators into $T$-periodic ones, and that it appears in the last term in Eq.~\eqref{evodiff}. Again we decompose it into its time average and the deviation from the average:
\begin{equation}
\mathcal{K}_{t}(S_t^{(1)}) =  \overline{\mathcal{K}_{t}(S_t^{(1)})} + \Delta (\mathcal{K}_{t}(S_t^{(1)})). 
\end{equation}
We can then perform another integration by parts for the deviation part. Iterating this procedure $L$ times, we obtain the following expression for the difference between the two evolutions:
\begin{widetext}
\begin{equation}
U_{1}(t)-U_{2}(t) 
=\sum_{k=1}^{L+1}(-i)^kS_t^{(k)}U_{2}(t)
-i\int_{0}^{t}ds\,U_{1}(t) U_{1}^{\dagger}(s)\,\Bigg(\sum_{k=0}^{L} (-i)^k\overline{\mathcal{K}_{s}(S_s^{(k)})} + (-i)^{L+1}\mathcal{K}_{s}(S_{s}^{(L+1)}) \Bigg)\,U_{2}(s),
\label{generror}
\end{equation}
\end{widetext}
where the functions ${S}^{(k)}_t$ defined by
\begin{equation}
\begin{cases}
\medskip
\displaystyle
S_{t}^{(0)} =\openone,&\\
\displaystyle
S_{t}^{(k)} =\int_{0}^{t}ds\,\Bigl(\mathcal{K}_{s}(S_{s}^{(k-1)})-\overline{\mathcal{K}_{s}(S_{s}^{(k-1)})}\Bigr)&(k\ge1)
\end{cases}
\label{defthree}
\end{equation}
play the role of $k$th-order integral actions.

Here and in the following, we shall denote by $\Omega=2\pi/T$ the frequency of the two Hamiltonians. With this notation, one readily sees that the integral actions satisfy
\begin{equation}\label{eq:actions_order}
S_t^{(k)}=\mathcal{O}(\Omega^{-k})
\end{equation}
at any given time $t$. We also note that each of these functions, except the $0$th one, vanishes at the full period:
\begin{equation}\label{eqn:VanishingActions}
S_T^{(k)}=0\quad(k\ge 1).
\end{equation}

\subsection{Deriving Effective Hamiltonians}
Equation~\eqref{generror} can be directly used to bound the distance between the evolutions induced by two arbitrary periodic Hamiltonians $H_1(t)$ and $H_2(t)$ sharing a common period, irrespective of whether this distance is small or not. Here we will take one step forward and utilize this formula to \textit{construct} effective Hamiltonians: given a periodic Hamiltonian $H(t)$ with period $T$, we will search for a time-independent Hamiltonian in the form
\begin{equation}
H_{\mathrm{eff},L} = \sum_{k=0}^{L} \frac{1}{\Omega^k}H_\mathrm{eff}^{(k)},
\label{eqn:Heff}
\end{equation}
again with $\Omega=2\pi/T$, such that the evolutions generated by $H(t)$ and $H_{\mathrm{eff},L}$ are close up to the desired precision. This will be done by using Eq.~\eqref{generror} with $H_1(t)=H(t)$, $H_2(t)=H_{\mathrm{eff},L}$, and fixing the Hermitian operators $H_\mathrm{eff}^{(k)}$ ($k=1,\ldots,L$) so that
\begin{equation}
\sum_{k=0}^{L}(-i)^k\overline{\mathcal{K}_{s}(S_s^{(k)})}=\mathcal{O}(\Omega^{-(L+1)})
\label{eqn:CondEff}
\end{equation}
is satisfied in the expansion~\eqref{generror}. Indeed, if this condition is satisfied, by applying Eq.~\eqref{generror} and taking Eq.~\eqref{eq:actions_order} into account, the distance between the evolutions satisfies 
\begin{equation}\label{eq:approx}
    \|U(t)-e^{-i H_{\mathrm{eff},L}t}\|=\mathcal{O}(\Omega^{-1})
\end{equation}
up to times of order $t\sim\mathcal{O}(\Omega^L)$. 
Above, as well as in the remainder of the paper, $\|\cdot\|$ shall denote the \textit{operator norm} on $\mathcal{H}$, defined by $\|T\|=\sup_{0\neq x\in\mathcal{H}}\|Tx\|/\|x\|$ for every bounded linear operator $T$.
In other words, the condition~\eqref{eqn:CondEff} is sufficient to define an effective time-independent Hamiltonian whose dynamics approximates the real one up to times of order $t\sim\mathcal{O}(\Omega^L)$.
We can thus extend the validity of the approximation longer and longer as we collect higher- and higher-order corrections to the effective Hamiltonian $H_\mathrm{eff}$.

We also notice that, under the same condition~\eqref{eqn:CondEff}, the distance between the evolutions induced by $H(t)$ and $H_{\rm eff}$ becomes especially good after a full period. Indeed, by using again Eq.~\eqref{generror} with $H_1(t)=H(t)$, $H_2(t)=H_{\rm eff}$, inserting the condition~\eqref{eqn:CondEff}, and taking Eq.~\eqref{eqn:VanishingActions} into account, we get
\begin{equation}\label{eq:approx_period}
    \|U(T)-e^{-iH_{\mathrm{eff},L}T}\|=\mathcal{O}(\Omega^{-(L+1)}).
\end{equation}
Therefore, our technique to derive effective Hamiltonians is tailored to approximate the real dynamics with arbitrarily high precision over full periods.

\subsection{Equivalence to the Floquet--Magnus Expansion}
The Floquet--Magnus expansion is a systematic approach for deriving effective Hamiltonians in periodically driven systems~\cite{Eckardthigh,Bukovuniversal,rahaveff,Casasfloquet,goldmanperiod,goldmandriven}. We shall briefly recall this approach here. In the Floquet theory, given a time-dependent Hamiltonian $H(t)$ with period $T$ and an initial time $t_0$, one defines the so-called Floquet Hamiltonian $H_{\rm F}[t_0]$ via the following equality:
\begin{equation}
U(t_{0}+T, t_{0}) =\mathcal{T} \exp\!\left(-i\int_{t_{0}}^{t_{0}+T}dt\,H(t)\right) = e^{-iH_{\rm F}[t_{0}]T},
\end{equation}
with $U(t,t_0)$ being the unitary evolution generated by $H(t)$. This equality uniquely defines $H_{\rm F}[t_0]$ up to a branch choice for the logarithm. With this definition, the unitary evolution of the system can be decomposed as
\begin{equation}\label{eq:evol_decomposed}
U(t,t_{0}) = U_\mathrm{F}(t,t_0)e^{-iH_{\rm F}[t_0](t-t_{0})}.
\end{equation}
The argument $[t_0]$ indicates that the Floquet Hamiltonian $H_{\rm F}[t_0]$ depends on the initial time $t_0$ (see Ref.~\cite{Eckardthigh} for elaborations on this point). 
The Floquet Hamiltonian $H_{\rm F}[t_0]$ captures the evolution of the driven system at stroboscopic times $t_0+mT$, $m=0,1,2,\ldots$ (macromotion), while the evolution between the stroboscopic times (micromotion) is described by $U_\mathrm{F}(t,t_0)$~\cite{Eckardthigh}, which is periodic in both arguments, $U_\mathrm{F}(t+T, t_0) = U_\mathrm{F}(t, t_0+T) = U_\mathrm{F}(t, t_0)$.

In general, it is hard to derive the Floquet Hamiltonian $H_{\rm F}[t_0]$ exactly. The Magnus expansion~\cite{magnuson, Blanespeda, BLANESmagnus} provides a useful approach to systematically compute the Floquet Hamiltonian $H_{\rm F}[t_0]$, particularly in the high-frequency regime, where the convergence of the Magnus expansion is ensured. The resulting expansion of $H_{\rm F}[t_0]$, the  Floquet--Magnus expansion, reads
\begin{equation}
H_{\rm F}[t_{0}] = \sum_{k=0}^{\infty} H^{(k)}_\mathrm{F}[t_{0}],
\label{eqn:ME}
\end{equation}
with the first two terms given by the formulae
\begin{align}
H_{\rm F}^{(0)}[t_0] &= \frac{1}{T} \int_{t_{0}}^{t_{0}+T}dt\,H(t),
\label{eqn:MS0}\\
H_{\rm F}^{(1)}[t_0] &= \frac{1}{2iT} \int_{t_{0}}^{t_{0}+T}dt_{1} \int_{t_{0}}^{t_{1}} dt_{2}\,[H(t_{1}), H(t_{2})].
\label{eqn:MS1}
\end{align}
Higher-order terms involve nested commutators of the Hamiltonian $H(t)$ and can be computed systematically~\cite{BLANESmagnus}. The full Floquet--Magnus expansion also provides a series for the micromotion $U_\mathrm{F}(t,t_0)$, but our focus is the macromotion, for which the Floquet--Magnus series and the Magnus series at full period coincide up to a factor of $T$~\cite{Casasfloquet}, in the following sense. Recall that, in the Magnus expansion, one writes the unitary evolution as  
\begin{align}
U(t,t_0) &= e^{A(t,t_0)}, \quad 
A(t,t_0) = \sum_{n=1}^\infty A_n(t,t_0).
\end{align}
Therefore, at stroboscopic times,  
\begin{align}
U(t_0+T,t_0) &= e^{A(T)} = e^{-i H_\mathrm{F}[t_0]T},
\end{align}
so that $A(T) \equiv A(t_0+T,t_0)= -i H_\mathrm{F}[t_0]T$ with a proper choice of logarithmic branch. Hence, at a full period, the Magnus and Floquet--Magnus expansions are equivalent, differing only by the overall factor $T$.

We claim that, for every integer $L$, the effective Hamiltonian $H_{\mathrm{eff},L}$ derived in the previous subsection, cf.~Eq.~\eqref{eqn:Heff}, exactly corresponds to the sum of the first $L$ terms in the Floquet--Magnus expansion~\eqref{eqn:ME} of $H_{\rm F}[t_0]$. In other words, the two methods produce the same effective Hamiltonians at any order. 
Without loss of generality, we will prove this claim for $t_0=0$, and set $U(t,0)\equiv U(t)$. 
First notice that, for every $k\leq L$, the $k$th-order term in the Floquet--Magnus expansion~\eqref{eqn:ME} satisfies
\begin{equation}\label{eq:orda}
    H_{\rm F}^{(k)}[0]=\mathcal{O}(\Omega^{-k}),
\end{equation}
just as the $k$th-order term in the expansion~\eqref{eqn:Heff} of our effective Hamiltonian $H_{\mathrm{eff},L}$. Equation~\eqref{eq:orda} implies 
\begin{equation}
    \left\|H_{\rm F}[0]-H_{\mathrm{F},L}[0]\right\|=\mathcal{O}(\Omega^{-(L+1)}),
\end{equation}
where $H_{\mathrm{F},L}[0]$ is the sum of the first $L$ terms in the Floquet--Magnus expansion~\eqref{eqn:ME}. Therefore, the difference between the corresponding evolutions at $t=T$ is of the same order:
\begin{equation}\label{eq:approx_period_FM}
    \|U(T)-e^{-iH_{\mathrm{F},L}[0]T}\|=\mathcal{O}(\Omega^{-(L+1)}),
\end{equation}
where we used the fact that, at each full period, the evolution induced by the Floquet Hamiltonian reproduces the exact one [cf.~Eq.~\eqref{eq:evol_decomposed}]. The same property holds for the evolution induced by our $L$th-order effective Hamiltonian [cf.~Eq.~\eqref{eq:approx_period}]. Combining Eqs.~\eqref{eq:approx_period} and~\eqref{eq:approx_period_FM} we derive
\begin{equation}\label{eq:approx_period_FM2}
    \|e^{-iH_{\mathrm{eff},L}T}-e^{-iH_{\mathrm{F},L}[0]T}\|=\mathcal{O}(\Omega^{-(L+1)}).
\end{equation}

As proven in Appendix~\ref{app:bound}, for $\Omega$ large enough there exists a constant $C>0$ such that
\begin{equation}\label{eq:approx_period_FM3}
    \|H_{\mathrm{eff},L}-H_{\mathrm{F},L}[0]\|\leq C\|e^{-iH_{\mathrm{eff},L}T}-e^{-iH_{\mathrm{F},L}[0]T}\|,
\end{equation}
thus implying
\begin{equation}\label{eq:approx_period_FM4}
    \|H_{\mathrm{eff},L}-H_{\mathrm{F},L}[0]\|=\mathcal{O}(\Omega^{-(L+1)}).
\end{equation}
As both $H_{\mathrm{eff},L}$ and $H_{\mathrm{F},L}[0]$ are polynomials in $\Omega^{-1}$ of order $L$, this implies $H_{\mathrm{eff},L}=H_{\mathrm{F},L}[0]$. Therefore, the two Hamiltonians exactly coincide order by order, as claimed.

As such, the Floquet--Magnus expansion and our strategy to derive effective Hamiltonians yield the same result. Remarkably, however, our method also provides an explicit error bound on the distance between the evolutions generated by the original and effective Hamiltonians. Equivalently, our method provides rigorous, quantitative error bounds for the Floquet--Magnus expansion---independently of whether said expansion converges.

\section{Application to the Semiclassical Rabi Model and Explicit Error Bounds}
\label{BS}
We now apply our formalism to the semiclassical Rabi Hamiltonian $H(t)$ in Eq.~\eqref{rabiham} with a large driving frequency $\omega$.
To this end, we begin as usual by transforming to the rotating frame with respect to the free Hamiltonian $H_0=(\omega/2)Z$ of the Hamiltonian $H(t)$.
In this frame, the Hamiltonian reads
\begin{align}
\hat{H}(t) &= e^{iH_0t}
[H(t)-H_0]
e^{-iH_0t}
\nonumber\\
&=\frac{\delta}{2}Z 
+ \frac{g}{2}X
+ \frac{g}{2}[\cos(2\omega t)X-\sin(2\omega t)Y],
\end{align}
where $\delta$ is the detuning between the natural and driving frequencies of the system: $\delta=\omega_0 - \omega$.
Hereafter we focus on the resonant case $\delta=0$, i.e.~$\omega = \omega_{0}$, whence
\begin{equation}
\hat{H}(t)
=\frac{g}{2}X+\frac{g}{2}[\cos(2\omega t)X-\sin(2\omega t)Y].
\label{reshamiltonian}
\end{equation}
We point out that, as this Hamiltonian oscillates with frequency $2\omega$, the quantity $\Omega$ as introduced in Section~\ref{general iteration} corresponds to $\Omega=2\omega$.

Before applying our formalism to this Hamiltonian, let us compute the first two terms of its Floquet--Magnus expansion~(\ref{eqn:ME}) via the formulae~\eqref{eqn:MS0} and~\eqref{eqn:MS1} with $T=2\pi/\Omega=\pi/\omega$. We take $t_{0}=0$ and omit the explicit indication of the initial time, that is, $H_{\rm F}^{(n)}\equiv H_{\rm F}^{(n)}[0]$. The first contribution reads
\begin{equation}
H_{\rm F}^{(0)} =\frac{g}{2}X.
\end{equation}
This corresponds to the standard RWA, which captures the time-averaged part of the Hamiltonian.
The second contribution reads
\begin{equation}
H_{\rm F}^{(1)}
=-\frac{g^{2}}{8\omega} Z.
\end{equation}
This is the well-known Bloch--Siegert shift~\cite{BlochMag}.

We will now apply the iterative strategy introduced in Section~\ref{general iteration} to the semiclassical Rabi model. As proven in the general case, our procedure will reproduce the same results of the Floquet--Magnus expansion, but will also yield an explicit bound on the corresponding error on the dynamics.

\subsection{Derivation of the Bloch--Siegert Hamiltonian} 
We aim to find a time-independent effective Hamiltonian $H_\mathrm{eff}$ approximating the evolution generated by the Hamiltonian $\hat{H}(t)$ in the rotating frame up to order $\mathcal{O}(\omega^{-1})$. To this end, we use the formula~(\ref{generror}) for $H_1(t)=\hat{H}(t)$ and $H_2(t)=H_\mathrm{eff}$, and consider the expansion
\begin{equation}
H_\mathrm{eff,1}=H_\mathrm{eff}^{(0)}+\frac{1}{\omega}H_\mathrm{eff}^{(1)}.
\end{equation}
We will determine the operators
$H_\mathrm{eff}^{(0)}$ and $H_\mathrm{eff}^{(1)}$ by imposing the condition~\eqref{eqn:CondEff} for $L=1$ to be satisfied. We list explicit expressions for all relevant quantities in the formula~\eqref{generror}. Direct computations yield
\begin{align}
\mathcal{K}_{t}(S_{t}^{(0)}) 
={}&\hat{H}(t)-H_\mathrm{eff,1}\nonumber\\
={}&\frac{g}{2}X+\frac{g}{2}[\cos(2\omega t)X-\sin(2\omega t)Y]\nonumber\\
&
{}-H_\mathrm{eff}^{(0)}-\frac{1}{\omega}H_\mathrm{eff}^{(1)},
\\
S_{t}^{(1)}(t)
={}&\int_{0}^{t}ds\left(\mathcal{K}_{s}(S_{s}^{(0)})-\overline{\mathcal{K}_{s}(S_{s}^{(0)})}\right)
\nonumber\\
={}&\frac{g}{4\omega}\,\Bigl(\sin(2\omega t)X-[1-\cos(2\omega t)]Y\Bigr),
\\
\mathcal{K}_{t}(S_{t}^{(1)}(t))
&=\hat{H}(t)S_{t}^{(1)}(t)-S_{t}^{(1)}(t)H_\mathrm{eff,1}
\nonumber\\
={}&\frac{g^2}{4\omega}
\sin(2\omega t)\openone
\nonumber\\
&
{}-\frac{g}{4\omega^2}\,\Bigl(\sin(2\omega t)X-[1-\cos(2\omega t)]Y\Bigr)\,
H_\mathrm{eff}^{(1)},
\\
S_{t}^{(2)}(t)
={}&\int_{0}^{t}ds\left(\mathcal{K}_{s}(S_{s}^{(1)})-\overline{\mathcal{K}_{s}(S_{s}^{(1)})}\right)\nonumber\\
={}&\frac{g^2}{8\omega^2}
[1-\cos(2\omega t)]\openone
\nonumber\\
&
{}-\frac{g}{8\omega^2}\,\Bigl([1-\cos(2\omega t)]X+\sin(2\omega t)Y\Bigr)\,
H_\mathrm{eff}^{(0)}
\nonumber\\
&
{}-\frac{g}{8\omega^3}\,\Bigl([1-\cos(2\omega t)]X+\sin(2\omega t)Y\Bigr)\,
H_\mathrm{eff}^{(1)},
\end{align}
With these quantities in hand, we thus obtain
\begin{equation}
\overline{\mathcal{K}_{s}(S_{s}^{(0)})}=\frac{g}{2}X-H_\mathrm{eff}^{(0)}-\frac{1}{\omega}H_\mathrm{eff}^{(1)}.
\end{equation}
Our general condition~\eqref{eqn:CondEff} reads $\overline{\mathcal{K}_{s}(S_{s}^{(0)})}=\mathcal{O}(\omega^{-1})$,
which we achieve by setting
\begin{equation}
H_\mathrm{eff}^{(0)}=\frac{g}{2}X.
\end{equation}
As expected, this is the standard RWA for the semiclassical model.
To derive the higher-order effective model, we likewise look at the next order $\omega^{-1}$ and impose it to vanish as well. We get contributions from the first two terms
\begin{align}
&
\overline{\mathcal{K}_{s}(S_{s}^{(0)})}
-i\overline{\mathcal{K}_{s}(S_{s}^{(1)})}
\nonumber\\
&\quad
=\frac{g}{2}X-H_\mathrm{eff}^{(0)}-\frac{1}{\omega}H_\mathrm{eff}^{(1)}
-\frac{ig}{4\omega}YH_\mathrm{eff}^{(0)}
-\frac{ig}{4\omega^2}YH_\mathrm{eff}^{(1)}.
\end{align}
By using $H_\mathrm{eff}^{(0)}=\frac{g}{2}X$, the condition~\eqref{eqn:CondEff} for $L=1$ yields
\begin{equation}
H_\mathrm{eff}^{(1)}
=-\frac{g^2}{8}Z.
\end{equation}
To summarize, at the first two iterative orders, our techniques precisely yields the RWA Hamiltonian and the Bloch--Siegert Hamiltonian:
\begin{equation}
H_{\rm RWA}=H_\mathrm{eff,0}=\frac{g}{2}X
\end{equation}
and
\begin{equation}
H_{\rm BS}=H_\mathrm{eff,1}=\frac{g}{2}X-\frac{g^{2}}{8\omega}Z,
\end{equation}
compatibly, as expected, with the Floquet--Magnus expansion.

\subsection{Error Bound between the Evolutions}
\label{error bound}
As already mentioned, our technique allows us to directly establish analytic error bounds to the validity of the Bloch--Siegert approximation, generalizing the results in Ref.~\cite{Onebound} where such a bound was provided for the RWA\@,

\begin{equation}
\Vert U(t)-e^{-iH_{\rm RWA}t}\Vert \leq \frac{g}{2\omega}\left(1+ \frac{g}{2}t\right).
\label{ERWA}
\end{equation}

This is of crucial importance  for determining the validity of the approximation, as it provides insight how an evolution generated by an effective Hamiltonian diverges from the original evolution over the time and, importantly, clarifies the parameter regime within which the approximation holds.

To this purpose, we consider Eq.~\eqref{generror} with $H_1(t)=\hat{H}(t)$, $H_2(t)=H_{\rm BS}$, and $L=1$. We take norms on both sides of Eq.~\eqref{generror} and bound the distance between the evolutions by combining $\sup_t\|\sum_{k=1}^{L+1}(-i)^kS_t^{(k)}\|$,
$\|\sum_{k=0}^{L} (-i)^k\overline{\mathcal{K}_{s}(S_s^{(k)})}\|$, and $\sup_s\|\mathcal{K}_{s}(S_{s}^{(L+1)})\|$ via the triangle inequality. 
We get the following bound in the case of resonant driving ($\omega=\omega_0$):

\begin{align}
&
\Vert U(t)-e^{-iH_{\rm BS}t}\Vert 
\nonumber\\
&
\quad\leq\frac{g}{2\omega}\sqrt{   1+\frac{3g^2}{16\omega^2} +\frac{g^4}{256\omega^4}}
+\frac{3g^3t}{32\omega^2}\left(
1+\frac{g^2}{24\omega^2}
\right).
\label{EB}
\end{align}

% Throughout this work, we use the operator norm when deriving bounds. The operator norm is defined as follows: let $X$ and $Y$ be normed spaces, and $T : X \rightarrow Y$ a bounded linear operator. The operator norm is
% \begin{equation}
% \|T\| = \sup_{x \in X,\, x \neq 0} \frac{\|T x\|_{Y}}{\|x\|_{X}}.
% \end{equation}

In Appendix~\ref{mathematica}, we provide a Mathematica script which automatically computes the effective Hamiltonian $H_{\mathrm{eff},L}$ and the quantities relevant for the estimation of the distance. As expected, the error bound goes to zero as $\omega \rightarrow \infty$, thus rigorously certifying the validity of the Bloch--Siegert approximation. Similarly to the bound for the RWA derived in Ref.~\cite{Onebound}, this bound consists of a time-independent term, which dominates for small $t$, and a term which increases linearly with $t$, eventually rendering the bound trivial at sufficiently large times. In both regimes, the bound outperforms the one for the RWA both as $g\to0$ and as $\omega\to\infty$, as both parameters appear in Eq.~\eqref{EB} in higher powers.

A bound for nonresonant driving with $\delta=\omega_0-\omega$ can also be derived using our method:
\begin{widetext}
\begin{align}
&
\Vert U(t)-e^{-iH_{\rm BS}t}\Vert 
\nonumber\\
&\quad
\leq
\frac{g}{2 \omega}
\sqrt{
 \left(1-\frac{\delta}{2\omega}\right)^2
+ \frac{3g^2}{16\omega^2} \left(
1- \frac{2\delta}{3\omega} 
+\frac{\delta^2}{12\omega^2} 
\right)
+\frac{g^4}{256\omega^4} 
}
+ \frac{g^3t}{32\omega^2}\sqrt{1+\frac{4\delta^2}{g^2}}
\nonumber \\
&\quad\quad{}
+ \frac{g^3t}{16\omega^2} \sqrt{
 9
    + \frac{4\delta}{\omega}
 +     \frac{4\delta^2}{\omega^2}
 + \frac{\delta^3}{2\omega^3}
 +\frac{\delta^4}{16\omega^4} 
+  \frac{g^2}{8\omega^2} \left(
5 + \frac{\delta}{\omega} +\frac{\delta^2}{4\omega^2}
\right) 
+\frac{g^4}{256\omega^4} 
+ 
 \frac{4\delta^2}{g^2} \left(
     7
+ \frac{4\delta^2}{g^2} 
 + \frac{\delta}{\omega} 
+ \frac{3\delta^2}{4\omega^2} 
    \right) 
}.
\label{EB_nonreso}
\end{align}
\end{widetext}
This bound is valid for $-1<\delta/\omega\le2 (5 - 2 \sqrt{5})$.

Note that we always have $\delta/\omega=\omega_0/\omega-1>-1$ since $\omega_0,\omega>0$.
We also note that this bound does not reduce to the bound~(\ref{EB}) in the limit $\delta\to0$. This is because we employed an upper bound on $\sup_s\|\mathcal{K}_{s}(S_{s}^{(L+1)})\|$ to get the bound~(\ref{EB_nonreso}) while we estimated $\sup_s\|\mathcal{K}_{s}(S_{s}^{(L+1)})\|$ precisely to get the bound~(\ref{EB}).

\begin{figure}[b]
\begin{center}
\begin{minipage}{\linewidth}
\includegraphics[width=\textwidth]{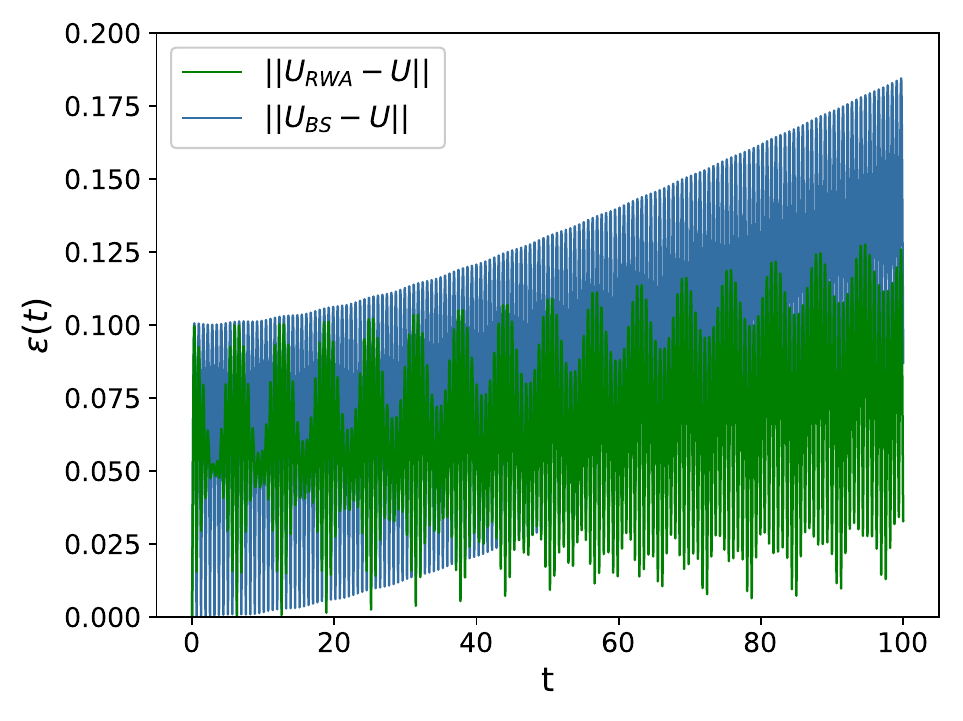}
\caption{Numerical distance (in the operator norm) between the exact evolution generated by the semiclassical Rabi Hamiltonian, and the evolutions of the RWA (orange) and Bloch--Siegert (blue) Hamiltonians, for large times. We set $g=1$ and $\omega = 5$ for the numerics.}
\label{RWABS}
\end{minipage}
\end{center}
\end{figure}
\begin{figure}[b]
\begin{center}
\begin{minipage}{\linewidth}
\includegraphics[width=\textwidth]{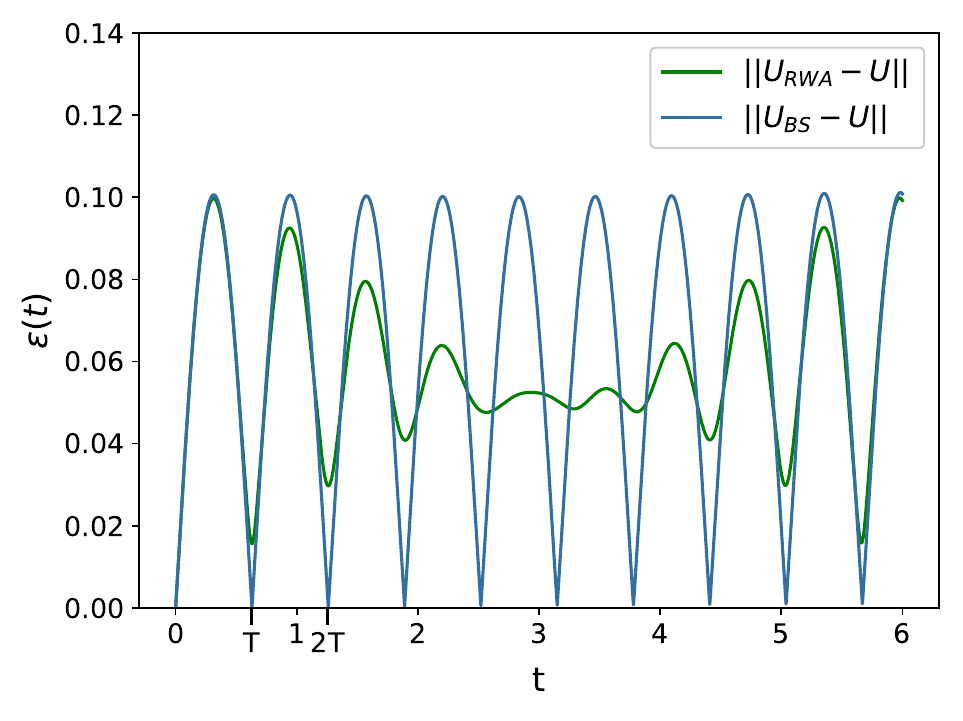}
\caption{Numerical distance (in the operator norm) between the exact evolution generated by the semiclassical Rabi Hamiltonian, and the evolutions of the RWA (orange) and Bloch--Siegert (blue) Hamiltonians, for short times. We set $g=1$ and $\omega = 5$ for the numerics.}
\label{period}
\end{minipage}
\end{center}
\end{figure}

\subsection{Numerical Comparison between Bloch--Siegert and RWA} 
We now proceed to compare the evolutions induced by the semiclassical Rabi model and its Bloch--Siegert approximation numerically. In general, higher-order effective Hamiltonians are expected to provide more accurate approximations by capturing finer details of the original interaction. Accordingly, one would expect that the Bloch--Siegert will generate a closer dynamics to the exact evolution than the RWA. In fact, as shown in Fig.~\ref{RWABS}, this is not the case: the RWA shows lower error than the Bloch--Siegert Hamiltonian, even over longer timescales.

It is insightful to examine the dynamics at stroboscopic times, i.e.~at integer multiples of the driving period, as the Bloch--Siegert Hamiltonian is constructed to closely approximate the exact evolution at those exact times [cf.~Eq.~\eqref{eq:approx_period}]. Indeed, as shown in Fig.~\ref{period}, the Bloch–-Siegert Hamiltonian outperforms the RWA at early stroboscopic times. However, as illustrated in Fig.~\ref{RWABS}, its accuracy deteriorates over longer times, and eventually it performs worse than the RWA even at those times. 

As we will see, in order to effectively improve the RWA, we will need to go to the third-order effective Hamiltonians, which will be shown to outperform the RWA both analytically \textit{and} numerically.

\section{Beyond the Second Order}\label{third order}
Previously, it was observed that the evolution governed by the Bloch--Siegert Hamiltonian does not exhibit improved dynamics compared to the RWA Hamiltonian, even over longer time scales. This result appears counter-intuitive. To investigate this further, we derive the third-order effective Hamiltonian and establish its bound in the resonant case, enabling a detailed comparison of its behavior with the RWA and Bloch--Siegert Hamiltonians.

\subsection{Third-Order Effective Hamiltonian}
As in the previous section, we use the iterative integration-by-parts method to derive the third-order Hamiltonian. Coming back to the resonant case $\delta=0$, we search for an effective Hamiltonian $H_\mathrm{eff,2}$ in the form
\begin{equation}
H_\mathrm{eff,2}=H_\mathrm{eff}^{(0)}+\frac{1}{\omega}H_\mathrm{eff}^{(1)}+\frac{1}{\omega^2}H_\mathrm{eff}^{(2)}.
\end{equation}
We first calculate the following quantities:
\begingroup
\allowdisplaybreaks
\begin{align}
\overline{\mathcal{K}_{t}(S_{t}^{(0)})}
={}&
\frac{g}{2}X-H_\mathrm{eff}^{(0)}-\frac{1}{\omega}H_\mathrm{eff}^{(1)}-\frac{1}{\omega^{2}}H_\mathrm{eff}^{(2)},
\\
\overline{\mathcal{K}_{t}(S_{t}^{(1)})} 
={}&
\frac{g}{4\omega}YH_\mathrm{eff}^{(0)}
+\frac{g}{4\omega^{2}}YH_\mathrm{eff}^{(1)}
+\frac{g}{4\omega^{3}}YH_\mathrm{eff}^{(2)},
\\
\overline{\mathcal{K}_{t}(S_{t}^{(2)})} 
={}&
\frac{g^{3}}{32\omega^{2}}X
\nonumber\\
&
{}-\frac{g^{2}}{8\omega^{2}}H_\mathrm{eff}^{(0)}
+\frac{g}{8\omega^{2}}XH_\mathrm{eff}^{(0)2}
-\frac{g^{2}}{8\omega^{3}}H_\mathrm{eff}^{(1)}
\nonumber\\
&
{}+\frac{g}{8\omega^{4}}XH_\mathrm{eff}^{(1)2}-\frac{g^{2}}{8\omega^{4}}H_\mathrm{eff}^{(2)}+\frac{g}{8\omega^{6}}XH_\mathrm{eff}^{(2)2}.
\end{align}
\endgroup
From the discussion in the previous section, we can identify the first two terms as  
$H_\mathrm{eff}^{(0)}= \frac{g}{2}X$ and $H_\mathrm{eff}^{(1)}=-\frac{g^2}{8}Z$.
To determine 
$H_\mathrm{eff}^{(2)}$, we impose the condition~\eqref{eqn:CondEff} for $L=2$. Since
\begin{equation}
\sum_{k=0}^2(-i)^k\overline{\mathcal{K}_{s}(S_s^{(k)})}
=
-\frac{1}{\omega^{2}}H_\mathrm{eff}^{(2)}
-\frac{g^3}{32\omega^{2}}X
+\mathcal{O}(\omega^{-3}),
\end{equation}
the condition~\eqref{eqn:CondEff} for $L=2$ requires
\begin{equation}
H_\mathrm{eff}^{(2)}=-\frac{g^3}{32}X.
\end{equation}
Collecting the three terms, we get
\begin{equation}
H_\mathrm{3rd}=H_\mathrm{eff,2} = \frac{g}{2}X-\frac{g^{2}}{8\omega}Z-\frac{g^{3}}{32 \omega^{2}}X.
\end{equation}
As expected, this coincides with the third-order approximation obtained by the Floquet-Magnus expansion. See for instance Ref.~\cite{ZEUCH}.

\subsection{Error Bound and Numerical Comparison between Different Orders}
Having identified $H_\mathrm{3rd}$, we can estimate a quantitative bound on the distance between the effective evolution and the exact one. To this purpose, we consider Eq.~\eqref{generror} with $H_1(t)=\hat{H}(t)$, $H_2(t)=H_\mathrm{3rd}$, and $L=2$. We take norms on both sides of Eq.~\eqref{generror} and bound the distance between the evolutions in the same way as done for the bound in Eq.~(\ref{EB}) to get
\begin{widetext}
\begingroup
\allowdisplaybreaks
\begin{align}
\Vert U(t)-e^{-iH_{\text{3rd}}t}\Vert 
\le{}&\frac{g}{2\omega}\sqrt{
1 
+ 5\left(\frac{g}{4\omega}\right)^2 
+ 4\left(\frac{g}{4\omega}\right)^4 
-  \left(\frac{g}{4\omega}\right)^6 
+  13\left(\frac{g}{4\omega}\right)^8
- 6\left(\frac{g}{4\omega}\right)^{10} 
+ \left(\frac{g}{4\omega}\right)^{12}
}
\nonumber\\
&{}+\frac{g^4t}{2(4\omega)^3}
\,\biggl(
\sqrt{
1
+\left(\frac{g}{4\omega}\right)^2
+2\left(\frac{g}{4\omega}\right)^4
+\left(\frac{g}{4\omega}\right)^6
}
\nonumber\\
&\hphantom{{}+\frac{g^4t}{2(4\omega)^3}\,\biggl(}
{}+2\sqrt{
1 
 - 3\left(\frac{g}{4\omega}\right)^2
 + 14\left(\frac{g}{4\omega}\right)^4
 -19\left(\frac{g}{4\omega}\right)^6
 +18\left(\frac{g}{4\omega}\right)^8
 - 7\left(\frac{g}{4\omega}\right)^{10}
 +\left(\frac{g}{4\omega}\right)^{12}
  }
\biggr).
\end{align}
\endgroup
\end{widetext}
This bound is valid for $g/\omega<2\sqrt{2}$.

\begin{figure}[b]
\begin{center}
\begin{minipage}{\linewidth}
\includegraphics[width=\textwidth]{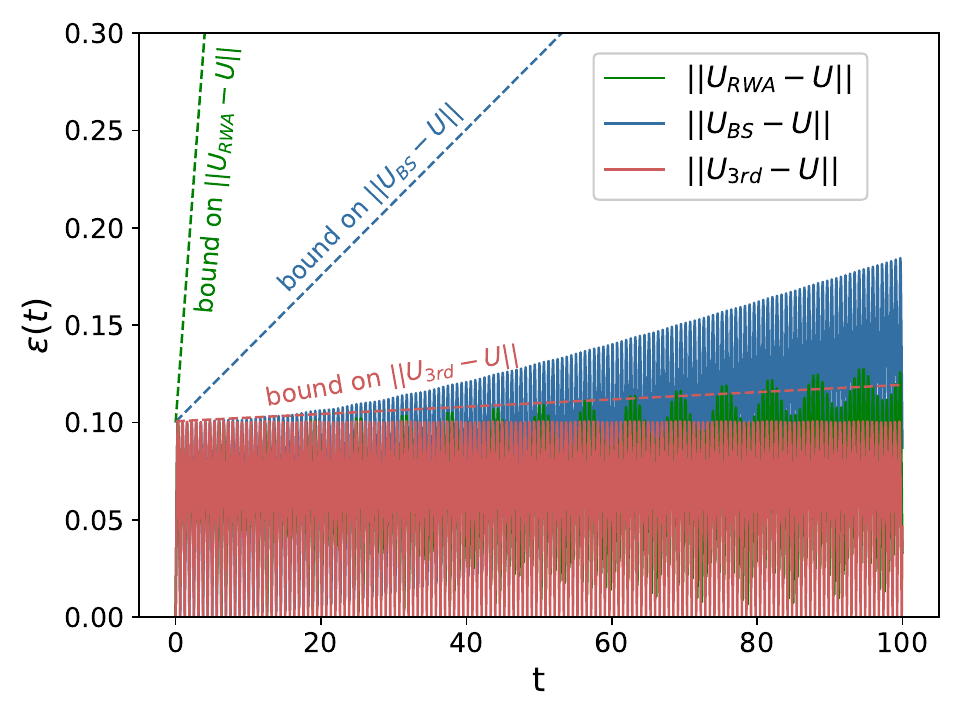}
\caption{
Numerical distance (in the operator norm) between the exact evolution generated by the semiclassical Rabi Hamiltonian, and the evolutions of the RWA (green), Bloch--Siegert (blue), and third-order (red) effective Hamiltonians, for large times. We set $g=1$ and $\omega = 5$ for the numerics. Additionally, the bounds for each of these effective Hamiltonians are shown as dashed lines.}
\label{comparison}
\end{minipage}
\end{center}
\end{figure}

\begin{figure}[b]
\begin{center}
\begin{minipage}{\linewidth}
\includegraphics[width=\textwidth]{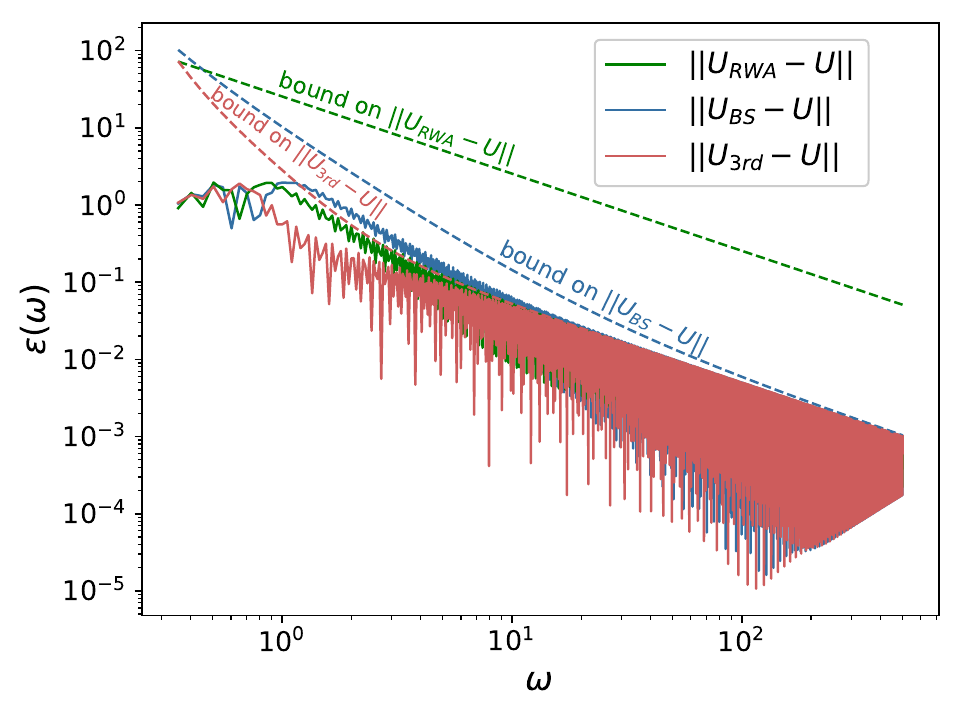}
\caption{
Log-log plot of the distance (in the operator norm) between the exact evolution generated by the semiclassical Rabi Hamiltonian, and the evolutions generated by the RWA (green), Bloch--Siegert (blue), and third-order (red) effective Hamiltonians, for large times. We set $g=1$ and $t = 100$ for the numerics. Additionally, the bounds for each of these effective Hamiltonians are shown as dashed lines.}
\label{comparison1}
\end{minipage}
\end{center}
\end{figure}

\begin{figure}[b]
\begin{center}
\begin{minipage}{\linewidth}
\includegraphics[width=\textwidth]{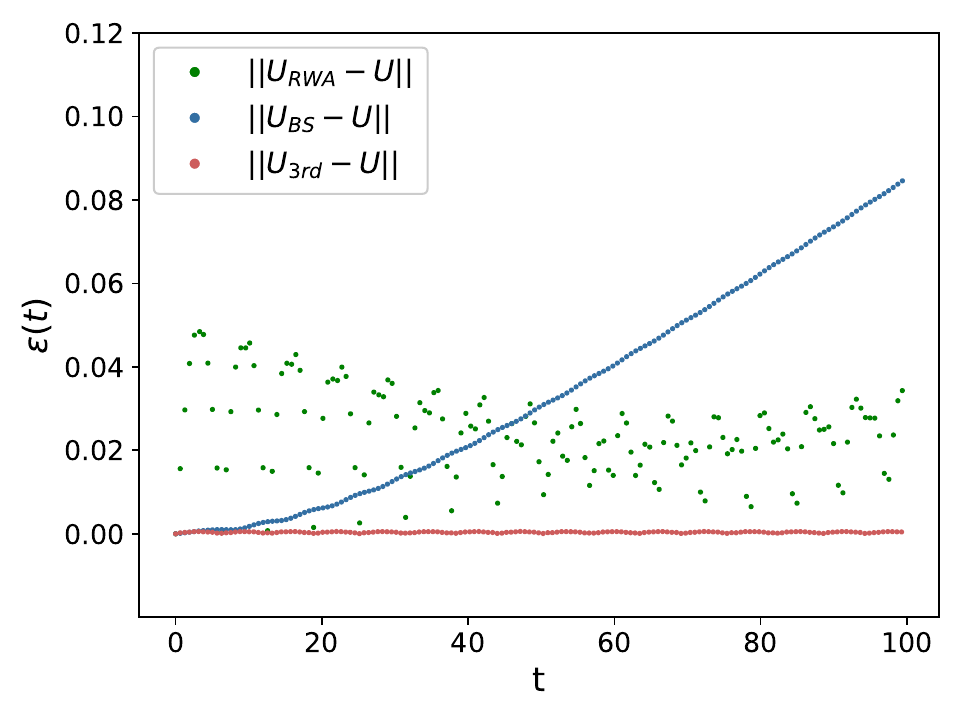}
\caption{Numerical distance (in the operator norm) between the exact evolution generated by the semiclassical Rabi Hamiltonian, and the evolutions of the RWA (green), Bloch--Siegert (blue), and third-order (red) effective Hamiltonians, for times multiple of the period. We set $g=1$ and $\omega = 5$ for the numerics.}
\label{multperiod}
\end{minipage}
\end{center}
\end{figure}

We complete this analysis by also providing a numerical comparison between the evolutions generated by the three effective Hamiltonians and the exact one. Fig.~\ref{comparison} and Fig.~\ref{comparison1} summarize our analytical and numerical results. While the bounds become loose over time, they reproduce the correct scaling in the frequency $\omega$ asymptotically. We can see that the dynamics generated by the third-order effective Hamiltonian $H_\mathrm{3rd}$ effectively outperforms both the RWA and Bloch--Siegert Hamiltonians. This is more clearly shown in Fig.~\ref{multperiod}, where we only evaluate the errors at multiples of the period. As discussed in the previous section, the Bloch--Siegert dynamics initially outperform the RWA at these times, but its validity eventually deteriorates. By contrast, the dynamics generated by $H_\mathrm{3rd}$ consistently outperforms both over very large timescales. This highlights the significance of the third-order effective Hamiltonian, which not only provides a tighter bound but also delivers more stable dynamics, with the error remaining better compared to the others for longer times.
This numerical analysis extends naturally to higher-order effective models. At the fourth order, simulations indicate that the dynamics closely tracks the third-order approximation for a substantial time window but eventually deviates more significantly. Therefore, the third-order model ultimately yields better long-time accuracy, much like the RWA outperforming the Bloch--Siegert Hamiltonian. This suggests the presence of a potential even–odd effect in the semiclassical Rabi model: the $(2k-1)$th-order approximations may maintain accuracy over longer timescales than their $(2k)$th-order counterparts. This is likely a model-specific effect not captured by our general bounds, thus highlighting the importance of combining analytical estimates with numerical insights to identify the most effective approximation schemes in practice.

\section{Conclusion}\label{conclusion}
In this paper we have developed a method based on the integration-by-parts lemma to derive nonperturbative effective Hamiltonians up to arbitrary order. Our method enables derivation of explicit error bounds between the original and the effective Hamiltonians. The method produces exactly the same effective Hamiltonians as of the Floquet--Magnus expansion, enabling us to find rigorous and quantitative bounds to its error at any order. Importantly, our bounds do not require the Magnus expansion to converge.

We applied our method to compute the Bloch--Siegert Hamiltonian from the Rabi Hamiltonian, established a rigorous and quantitative bound to the corresponding error in the dynamics, and also provided a numerical study of the performances of both approximations. Surprisingly, the latter analysis showed that over long times, the RWA Hamiltonian outperforms the Bloch--Siegert Hamiltonian, despite the latter being a higher-order approximation. Only for shorter times, the Bloch--Siegert performs better at the multiples of the driving period. 
To deepen our analysis, we derived the third-order effective Hamiltonian, computed its error bound, and compared its dynamics with the Bloch--Siegert and RWA Hamiltonians. Our analysis shows that the third-order Hamiltonian outperforms both and offers greater stability over longer times. Additionally, its error bound is significantly tighter, demonstrating more accuracy. 

This work presents a framework for deriving effective Hamiltonian for arbitrary orders and analyzing errors between effective and original evolutions across arbitrary times $t$ and coupling strengths $g$. Importantly, we showed that the error converges to zero as the driving frequency $\omega$ increases, confirming the validity of the approximations in the high-frequency regime. By explicitly deriving error bounds, our method offers a powerful tool for assessing the validity of effective Hamiltonians in diverse quantum systems. These findings are highly relevant for the advancement of future quantum technologies, where precise error quantification is critical for ensuring accuracy, scalability, and reliability.

Future research will be devoted to generalizing the framework presented here to unbounded Hamiltonians. Importantly, this would allow us to improve the results in Refs.~\cite{DBtaming,richter2024quantifying} by deriving higher-order effective Hamiltonians for the quantum Rabi model possibly outperforming the standard RWA\@.

\acknowledgments
D.L. and D.B. thank Robin Hillier for many stimulating discussions on the topic of the manuscript. A.D. was supported by Sydney quantum academy (SQA) through primary PhD scholarship. D.L. acknowledges financial support by Friedrich-Alexander-Universit\"at Erlangen-N\"urnberg through the funding program ``Emerging Talent Initiative'' (ETI), and was partially supported by the project TEC-2024/COM-84 QUITEMAD-CM\@.
K.Y. acknowledges supports by the Top Global University Project from the Ministry of Education, Culture, Sports, Science and Technology (MEXT), Japan, and by JSPS KAKENHI Grant No.~JP24K06904 from the Japan Society for the Promotion of Science (JSPS)\@.

\appendix
\onecolumngrid

\section{Proof of Eq.~\eqref{eq:approx_period_FM3}}\label{app:bound}

In this appendix we provide a proof of Eq.~\eqref{eq:approx_period_FM3}. To this end, consider two operator-valued functions $\mathbb{R}\ni\lambda\mapsto A(\lambda),B(\lambda)$, with $A(\lambda)$ and $B(\lambda)$ being bounded self-adjoint operators for every $\lambda$. We also assume that the two functions $\lambda\mapsto A(\lambda),B(\lambda)$ are operator-norm continuous, and that $A(0)=B(0)=0$. Define $U_A(\lambda)=e^{iA(\lambda)}$ and $U_B(\lambda)=e^{iB(\lambda)}$. We claim that, for every $0<\epsilon<\pi$, there exists $C>0$ and $r>0$ such that, for every $|\lambda|<r$,
\begin{equation}\label{eq:ineq0}
    \|A(\lambda)-B(\lambda)\|\leq C\|U_A(\lambda)-U_B(\lambda)\|.
\end{equation}
The case considered in the main text is thus covered by setting $\lambda:=\Omega^{-1}$.

To prove Eq.~\eqref{eq:ineq0}, first note that, since $\lambda\mapsto A(\lambda),B(\lambda)$ are norm continuous and $A(0)=B(0)=0$, there exists $r>0$ such that, for every $|\lambda|<r$,
\begin{equation}
    \|A(\lambda)\|<\epsilon,\qquad\|B(\lambda)\|<\epsilon.
\end{equation}
In particular, the spectra of $A(\lambda)$ and $B(\lambda)$ are entirely contained in the segment $[-\epsilon,+\epsilon]$. Consequently, the spectra of $U_A(\lambda)$ and $U_B(\lambda)$ are contained inside the arch $\{e^{i\theta}:\theta\in[-\epsilon,+\epsilon]\}$. Furthermore, we have
\begin{equation}
    A(\lambda)=-i\log U_A(\lambda),\qquad B(\lambda)=-i\log U_B(\lambda),
\end{equation}
where $\log(\cdot)$ is the principal branch of the complex logarithm function, whose branch cut is on the negative real axis and thus does not intersect the spectra of the two unitaries $U_A(\lambda)$ and $U_B(\lambda)$. By using the Riesz--Dunford functional calculus (see e.g.~Ref.~\cite{Colombo2011}), we can write these logarithms as complex integrals:
\begin{equation}
    \log U_A(\lambda)
    =\frac{1}{2\pi i}\oint_{\Gamma}\log z\,[z-U_A(\lambda)]^{-1}\,dz,\qquad  \log U_B(\lambda)
    =\frac{1}{2\pi i}\oint_{\Gamma}\log z\,[z-U_B(\lambda)]^{-1}\,dz,
\end{equation}
where $\Gamma$ is any positively oriented closed curve that winds once around the arc $\{e^{i\theta}:\theta\in[-\epsilon,+\epsilon]\}$ containing the spectra of both unitaries $U_A(\lambda)$ and $U_B(\lambda)$, and stays away from the negative real axis. As such,
\begin{align}
    A(\lambda)-B(\lambda)
    =-i\log U_A(\lambda)+i\log U_B(\lambda)
    &=-\frac{1}{2\pi}\oint_\Gamma\log z\,\Bigl([z-U_A(\lambda)]^{-1}-[z-U_B(\lambda)]^{-1}\Bigr)\,dz\nonumber\\
    &=-\frac{1}{2\pi}\oint_\Gamma\log z\,[z-U_A(\lambda)]^{-1}[U_A(\lambda)-U_B(\lambda)][z-U_B(\lambda)]^{-1}\,dz,
\end{align}
where the second resolvent formula was used in the last step. Taking norms,
\begin{equation}\label{eq:ineq}
    \|A(\lambda)-B(\lambda)\|\leq\frac{1}{2\pi}\|U_A(\lambda)-U_B(\lambda)\|\oint_\Gamma|{\log z}|\|[z-U_A(\lambda)]^{-1}\|\|[z-U_B(\lambda)]^{-1}\||dz|.
\end{equation}
Let $d_\Gamma$ be the smallest value of the distance between the curve $\Gamma$ and the spectra of $U_A(\lambda)$ and $U_B(\lambda)$ (we chose $\Gamma$ so that $d_\Gamma>0$). Then, the resolvents of $U_A(\lambda)$ and $U_B(\lambda)$ satisfy
\begin{equation}
    \|[z-U_A(\lambda)]^{-1}\|\leq\frac{1}{d_\Gamma},\qquad\|[z-U_B(\lambda)]^{-1}\|\leq\frac{1}{d_\Gamma}.
\end{equation}
Furthermore, as $\Gamma$ is a finite path and $\log$ is continuous on it, $M_\Gamma\equiv\max_{x\in\Gamma}|{\log z}|<\infty$. Therefore, by Eq.~\eqref{eq:ineq} and the latter inequalities,
\begin{equation}
     \|A(\lambda)-B(\lambda)\|\leq\frac{M_\Gamma}{2\pi d_\Gamma^2}\left(\oint_\Gamma|dz|\right)\|U_A(\lambda)-U_B(\lambda)\|=\frac{M_\Gamma|\Gamma|}{2\pi d_\Gamma^2}\|U_A(\lambda)-U_B(\lambda)\|,
\end{equation}
where $|\Gamma|$ is the (finite) length of the path $\Gamma$, thus proving the claimed inequality with $C=M_\Gamma|\Gamma|/(2\pi d_\Gamma)^2$.

\section{Mathematica Script for Computing Effective Hamiltonians and Evaluation of Error Bounds}\label{mathematica}
We present here a Mathematica script implementing the method presented in this paper for deriving effective Hamiltonians. The script is capable of computing the terms of the effective Hamiltonian to arbitrary orders, and generates all necessary expressions for estimating error bounds between the dynamics governed by the effective and original Hamiltonians.

\subsection{Computation of $S_t^{(k)}$, $\mathcal{K}_{t}(S_{t}^{(k)})$, and $H_\mathrm{eff}^{(k)}$ terms}
The first part of the code computes the integrated action $S_t^{(k)}$, the functions $\mathcal{K}_t(S_{t}^{(k)})$, and the $H_\mathrm{eff}^{(k)}$ terms that constitute the effective Hamiltonians. 
\begin{lstlisting}[numbers=none]

(* Define the time-dependent periodic Hamiltonian here: H(t)=H(t+T) *)

H[t_] := ;

CC[t_] = -I H[t];

(* Define period T and \[CapitalOmega] in terms of the oscillation frequency from the Hamiltonian *)

T = ;
\[CapitalOmega] = 2 \[Pi]/T;

(* Generate basis matrices of arbitrary dimensions *)
Clear[GeneratePositionBasisMatrices]
GeneratePositionBasisMatrices[d_] := 
  Module[{basisMatrices, mat, i, j}, 
   basisMatrices = {}; 
   For[i = 1, i <= d, i++, 
    For[j = 1, j <= d, j++,
      mat = ConstantArray[0, {d, d}]; 
      mat[[i, j]] = 1;
      AppendTo[basisMatrices, mat]
    ];
   ];
   basisMatrices
  ];

(* Set dimension d *)
d = Length[CC[t]];
basisMatrices = GeneratePositionBasisMatrices[d];
Id = IdentityMatrix[d];

(* Set the order of approximation here. This also sets the number of iteration for the integration by parts *)
(* Set order of approximation *)
L = 1;
(* Set assumptions on variables g, \[Omega], t, \[CapitalOmega] *)
$Assumptions = {Array[v, d^2 - 1, Complex]};

(* Variables for iterative calculation *)
Clear[S, K, Kav, SumKav, Dterms]
(* Note that Dterms = -i H_{eff, L} *)
Dterms = {ConstantArray[0, {d, d}]};

(* Iterative procedure *)
For[l = 0, l <= L, l++, 
Print["\n"];
Print["* l = ", l];
AppendTo[Dterms, -I 1/\[CapitalOmega]^l (Array[v, d^2] . basisMatrices)];
Print["Dterms = ", Dterms];
DD = Sum[\[Lambda]^j Dterms[[j + 1]], {j, 0, l + 1}];
(* Initialize S and K *)
S = {Id};
K = {\[Lambda] CC[t] - DD};
Kav = {};
For[j = 0, j <= l, j++,
 KK[t_] = K[[j + 1]];
 (* AVERAGING: Express immediately in terms of \[CapitalOmega] *)
 AppendTo[Kav, (1/T) Integrate[KK[u], {u, 0, T}]];
 KKav = Kav[[j + 1]];
 (* Update S and K *)
 AppendTo[S, Integrate[KK[s] - KKav, {s, 0, t}]];
 AppendTo[K, \[Lambda] CC[t] . S[[j + 2]] - S[[j + 2]] . DD];
];
SumKav = Sum[Kav[[j + 1]], {j, 0, l}];
SeriesKav = Series[SumKav, {\[Lambda], 0, l + 1}] // FullSimplify;
 
Print["\!\(\*UnderoverscriptBox[\(\[Sum]\), \(j = 0\), \
\(l\)]\)\!\(\*OverscriptBox[SuperscriptBox[\(K\), \((j)\)], \
\(_\)]\) = ", SeriesKav];

Sol = Solve[
   Flatten[Table[SeriesCoefficient[SeriesKav[[i, j]], l + 1], {i, 1, d}, {j, 1, d}]] == Flatten[ConstantArray[0, {d, d}]], Array[v, d^2]];

Print["Sol = ", Sol];
Print["\n"];

Dterms = Dterms /. Flatten[Sol] // FullSimplify;

Print["\!\(\*SuperscriptBox[\(D\), \((j)\)]\) = ", Dterms];

Kav = Kav /. Flatten[Sol] // FullSimplify;

Print["\!\(\*OverscriptBox[SuperscriptBox[\(K\), \((j)\)], \(_\)]\) = ",
   Kav /. {\[Lambda] -> 1} // FullSimplify];

K = K /. Flatten[Sol] // FullSimplify;

Print["\!\(\*SuperscriptBox[\(K\), \((l + 1)\)]\) = ", 
   K[[(l + 1) + 1]] /. {\[Lambda] -> 1} // FullSimplify];

S = S /. Flatten[Sol] // FullSimplify;

Print["\!\(\*SuperscriptBox[\(S\), \((j)\)]\) = ", 
   S /. {\[Lambda] -> 1} // FullSimplify];

\end{lstlisting}

\subsection{Computation of the Error Bound}
In the next section of the code, the required norms of specific elements are computed to derive the explicit bound. In some cases Mathematica may struggle to simplify the solution or evaluate the norm of expressions such as $\sum_{k=1}^{L+1} S_t^{(k)}$ and $\sum_{k=0}^L\overline{\mathcal{K}_t(S_t^{(k)})} + \mathcal{K}_t(S_t^{(L+1)})$. When this occurs, we apply the triangle inequality to obtain a simpler but less tight bound.

\begin{lstlisting}[numbers=none]
(* Terms contributing to the error bound *)
Print["\n"];

ES = FullSimplify[\!\(
\*UnderoverscriptBox[\(\[Sum]\), \(j = 1\), \(l + 1\)]\(S[\([j + 1]\)]\)\)];

Print["\!\(\*UnderoverscriptBox[\(\[Sum]\), \(j = 1\), \(l + 1\)]\)\
\!\(\*SuperscriptBox[\(S\), \((j)\)]\)=", 
ES /. {\[Lambda] -> 1} // FullSimplify];
 
Print["=", Series[ES, {\[Lambda], 0, l + 2}] // FullSimplify];
 
ET1 = FullSimplify[\!\(
\*UnderoverscriptBox[\(\[Sum]\), \(j = 0\), \(l\)]\(Kav[[\(j + 1\)\(]\)]\)\)];

ET2 = FullSimplify[K[[(l + 1) + 1]]];

ET = FullSimplify[ET1 + ET2];

Print["\!\(\*UnderoverscriptBox[\(\[Sum]\), \(j = 0\), \(l\)]\)\
\!\(\*OverscriptBox[SuperscriptBox[\(K\), \((j)\)], \(_\)]\) + \
\!\(\*SuperscriptBox[\(K\), \((l + 1)\)]\)=", 
ET /. {\[Lambda] -> 1} // FullSimplify];

Print["=", Series[ET, {\[Lambda], 0, l + 2}] // FullSimplify];
 
Print["Norm[\!\(\*UnderoverscriptBox[\(\[Sum]\), \(j = 1\), \(l + 1\)]\)\
\!\(\*SuperscriptBox[\(S\), \((j)\)]\)]=", 
FullSimplify[Norm[ES /. {\[Lambda] -> 1}]]];
 
Print["Norm[\!\(\*UnderoverscriptBox[\(\[Sum]\), \(j = 0\), \(l\)]\)\
\!\(\*OverscriptBox[SuperscriptBox[\(K\), \((j)\)], \(_\)]\)]=",
FullSimplify[Norm[ET1 /. {\[Lambda] -> 1}]]];

Print["Norm[\!\(\*SuperscriptBox[\(K\), \((l + 1)\)]\)]=", 
FullSimplify[Norm[ET2 /. {\[Lambda] -> 1}]]];

Print["Norm[\!\(\*UnderoverscriptBox[\(\[Sum]\), \(j = 0\), \(l\)]\)\
\!\(\*OverscriptBox[SuperscriptBox[\(K\), \((j)\)], \(_\)]\) + \
\!\(\*SuperscriptBox[\(K\), \((l + 1)\)]\)]=", 
FullSimplify[Norm[ET /. {\[Lambda] -> 1}]]];
]
\end{lstlisting}

\twocolumngrid

\bibliographystyle{prsty-title-hyperref}
\bibliography{bsref}

\begin{thebibliography}{10}

\bibitem{BlochMag}
F. Bloch and A. Siegert, Magnetic Resonance for Nonrotating Fields,
  \href{https://doi.org/10.1103/PhysRev.57.522}{Phys. Rev. \textbf{57},  522
  (1940)}.

\bibitem{ShirleySol}
J.~H. Shirley, Solution of the {Schr\"odinger} Equation with a {Hamiltonian}
  Periodic in Time, \href{https://doi.org/10.1103/PhysRev.138.B979}{Phys. Rev.
  \textbf{138},  B979  (1965)}.

\bibitem{Mehring}
M. Mehring, \href{https://doi.org/10.1007/978-3-642-68756-3}{\textit{Principles
  of High Resolution {NMR} in Solids}, 2nd, revised and enlarged ed.}
  (Springer, Berlin, 1983).

\bibitem{GRIFONI}
M. Grifoni and P. H{\"{a}}nggi, Driven quantum tunneling,
  \href{https://doi.org/10.1016/S0370-1573(98)00022-2}{Phys. Rep. \textbf{304},
   229  (1998)}.

\bibitem{Lueffects}
Z. L{\"{u}} and H. Zheng, Effects of counter-rotating interaction on driven
  tunneling dynamics: Coherent destruction of tunneling and Bloch-Siegert
  shift, \href{https://doi.org/10.1103/PhysRevA.86.023831}{Phys. Rev. A
  \textbf{86},  023831  (2012)}.

\bibitem{Yanbloch}
Y. Yan, Z. L{\"{u}}, and H. Zheng, Bloch-Siegert shift of the Rabi model,
  \href{https://doi.org/10.1103/PhysRevA.91.053834}{Phys. Rev. A \textbf{91},
  053834  (2015)}.

\bibitem{Lauchtbreaking}
A. Laucht, S. Simmons, R. Kalra, G. Tosi, J.~P. Dehollain, J.~T. Muhonen, S.
  Freer, F.~E. Hudson, K.~M. Itoh, D.~N. Jamieson, J.~C. McCallum, A.~S.
  Dzurak, and A. Morello, Breaking the rotating wave approximation for a
  strongly driven dressed single-electron spin,
  \href{https://doi.org/10.1103/PhysRevB.94.161302}{Phys. Rev. B \textbf{94},
  161302  (2016)}.

\bibitem{Gameltime}
O. Gamel and D.~F.~V. James, Time-averaged quantum dynamics and the validity of
  the effective {Hamiltonian} model,
  \href{https://doi.org/10.1103/PhysRevA.82.052106}{Phys. Rev. A \textbf{82},
  052106  (2010)}.

\bibitem{ZEUCH}
D. Zeuch, F. Hassler, J.~J. Slim, and D.~P. DiVincenzo, Exact rotating wave
  approximation, \href{https://doi.org/10.1016/j.aop.2020.168327}{Ann. Phys.
  (N.Y.) \textbf{423},  168327  (2020)}.

\bibitem{Onebound}
D. Burgarth, P. Facchi, G. Gramegna, and K. Yuasa, One bound to rule them all:
  from {Adiabatic} to {Zeno},
  \href{https://doi.org/10.22331/q-2022-06-14-737}{Quantum \textbf{6},  737
  (2022)}.

\bibitem{DBtaming}
D. Burgarth, P. Facchi, R. Hillier, and M. Ligab{\`{o}}, Taming the Rotating
  Wave Approximation, \href{https://doi.org/10.22331/q-2024-02-21-1262}{Quantum
  \textbf{8},  1262  (2024)}.

\bibitem{richter2024quantifying}
L. Richter, D. Burgarth, and D. Lonigro, Quantifying the rotating-wave
  approximation of the {Dicke} model,
  \href{https://doi.org/10.48550/arXiv.2410.18694}{arXiv:2410.18694 [quant-ph]
  (2024)}.

\bibitem{hahn2024efficiency}
A. Hahn, D. Burgarth, and D. Lonigro, {Efficiency of dynamical decoupling for
  (almost) any spin–boson model},
  \href{https://scipost.org/10.21468/SciPostPhys.19.2.035}{SciPost Phys.
  \textbf{19},  035  (2025)}.

\bibitem{Rabion}
I.~I. Rabi, On the Process of Space Quantization,
  \href{https://doi.org/10.1103/PhysRev.49.324}{Phys. Rev. \textbf{49},  324
  (1936)}.

\bibitem{Rabispace}
I.~I. Rabi, Space Quantization in a Gyrating Magnetic Field,
  \href{https://doi.org/10.1103/PhysRev.51.652}{Phys. Rev. \textbf{51},  652
  (1937)}.

\bibitem{eberly}
L. Allen and J.~H. Eberly,
  \href{https://store.doverpublications.com/products/9780486655338?srsltid=AfmBOoqEJPcwUFBsnp5KTD05bsmuri__K7-p52uIrfydv5bdqoCs3Lwl}{\textit{Optical
  Resonance and Two-Level Atoms}} (Dover, New York, 1987).

\bibitem{cohenphoton}
C. Cohen‐Tannoudji, J. Dupont‐Roc, and G. Grynberg,
  \href{https://doi.org/10.1002/9783527618422}{\textit{Photons and Atoms:
  Introduction to Quantum Electrodynamics}} (Wiley, Weinheim, 1997).

\bibitem{Wangrotate}
P. Wang, E. Hiltunen, and J.~C. Schotland, Rotating wave approximation and
  renormalized perturbation theory,
  \href{https://doi.org/10.48550/arXiv.2311.02670}{arXiv:2311.02670 [quant-ph]
  (2023)}.

\bibitem{xie2010}
Q. Xie and W. Hai, Analytical results for a monochromatically driven two-level
  system, \href{https://doi.org/10.1103/PhysRevA.82.032117}{Phys. Rev. A
  \textbf{82},  032117  (2010)}.

\bibitem{Liang2024}
H. Liang, Generating arbitrary analytically solvable two-level systems,
  \href{https://doi.org/10.1088/1751-8121/ad26ab}{J. Phys. A: Math. Theor.
  \textbf{57},  095301  (2024)}.

\bibitem{liang2025}
H. Liang, S. Xia, Y. Chen, Y. Su, and J. Chen, Eigenvalue Spectra of {Rabi}
  Models with Infinite Matrix Representations,
  \href{https://doi.org/10.3390/axioms14040263}{Axioms \textbf{14},  263
  (2025)}.

\bibitem{ScullyZubairy}
M.~O. Scully and M.~S. Zubairy,
  \href{https://doi.org/10.1017/CBO9780511813993}{\textit{Quantum Optics}}
  (Cambridge University Press, Cambridge, 1997).

\bibitem{Giscarddynamics}
P.-L. Giscard and C. Bonhomme, Dynamics of quantum systems driven by
  time-varying {Hamiltonians}: Solution for the {Bloch-Siegert Hamiltonian} and
  applications to {NMR},
  \href{https://doi.org/10.1103/PhysRevResearch.2.023081}{Phys. Rev. Research
  \textbf{2},  023081  (2020)}.

\bibitem{pietobs}
I. Pietik\"ainen, S. Danilin, K.~S. Kumar, A. Veps\"al\"ainen, D.~S. Golubev,
  J. Tuorila, and G.~S. Paraoanu, Observation of the {Bloch-Siegert} shift in a
  driven quantum-to-classical transition,
  \href{https://doi.org/10.1103/PhysRevB.96.020501}{Phys. Rev. B \textbf{96},
  020501  (2017)}.

\bibitem{Zhangbloch}
J. Zhang, S. Saha, and D. Suter, Bloch-{Siegert} shift in a hybrid quantum
  register: Quantification and compensation,
  \href{https://doi.org/10.1103/PhysRevA.98.052354}{Phys. Rev. A \textbf{98},
  052354  (2018)}.

\bibitem{Saikobloch}
A.~P. Saiko, S.~A. Markevich, and R. Fedaruk, Bloch--{Siegert} oscillations in
  the {Rabi} model with an amplitude-modulated driving field,
  \href{https://doi.org/10.1088/1555-6611/ab4bc6}{Laser Phys. \textbf{29},
  124004  (2019)}.

\bibitem{Liexcitonic}
Y. Li, Y. Han, W. Liang, B. Zhang, Y. Li, Y. Liu, Y. Yang, K. Wu, and J. Zhu,
  Excitonic {Bloch-Siegert} shift in {$\mathrm{CsPbI}_3$} perovskite quantum
  dots, \href{https://doi.org/10.1038/s41467-022-33314-9}{Nat. Commun.
  \textbf{13},  5559  (2022)}.

\bibitem{slobosemi}
A.~O. Slobodeniuk, P. Koutensk\'y, M. Barto\ifmmode~\check{s}\else \v{s}\fi{},
  F. Troj\'anek, P. Mal\'y, T. Novotn\'y, and M. Koz\'ak, Semiconductor {Bloch}
  equation analysis of optical {Stark} and {Bloch-Siegert} shifts in monolayer
  ${\mathrm{WSe}}_{2}$ and ${\mathrm{MoS}}_{2}$,
  \href{https://doi.org/10.1103/PhysRevB.106.235304}{Phys. Rev. B \textbf{106},
   235304  (2022)}.

\bibitem{Lizuain2008}
I. Lizuain, J.~G. Muga, and J. Eschner, Vibrational Bloch-Siegert effect in
  trapped ions, \href{https://doi.org/10.1103/PhysRevA.77.053817}{Phys. Rev. A
  \textbf{77},  053817  (2008)}.

\bibitem{Warring2013}
U. Warring, C. Ospelkaus, Y. Colombe, K.~R. Brown, J.~M. Amini, M. Carsjens, D.
  Leibfried, and D.~J. Wineland, Techniques for microwave near-field quantum
  control of trapped ions,
  \href{https://doi.org/10.1103/PhysRevA.87.013437}{Phys. Rev. A \textbf{87},
  013437  (2013)}.

\bibitem{vedaie2023}
S.~S. Vedaie, E.~J. P{\'{a}}ez, N.~H. Nguyen, N.~M. Linke, and B.~C. Sanders,
  Bespoke pulse design for robust rapid two-qubit gates with trapped ions,
  \href{https://doi.org/10.1103/PhysRevResearch.5.023098}{Phys. Rev. Research
  \textbf{5},  023098  (2023)}.

\bibitem{Yudilevich_2023}
D. Yudilevich, A. Salhov, I. Schaefer, K. Herb, A. Retzker, and A. Finkler,
  Coherent manipulation of nuclear spins in the strong driving regime,
  \href{https://doi.org/10.1088/1367-2630/ad0c0b}{New J. Phys. \textbf{25},
  113042  (2023)}.

\bibitem{Series}
G.~W. Series, A semi-classical approach to radiation problems,
  \href{https://doi.org/10.1016/0370-1573(78)90070-4}{Phys. Rep. \textbf{43},
  1  (1978)}.

\bibitem{Viebahn}
K. Viebahn, Introduction to {Floquet} theory, 2020.

\bibitem{Casasfloquet}
F. Casas, J.~A. Oteo, and J. Ros, Floquet theory: exponential perturbative
  treatment, \href{https://doi.org/10.1088/0305-4470/34/16/305}{J. Phys. A:
  Math. Gen. \textbf{34},  3379  (2001)}.

\bibitem{rahaveff}
S. Rahav, I. Gilary, and S. Fishman, Effective {Hamiltonians} for periodically
  driven systems, \href{https://doi.org/10.1103/PhysRevA.68.013820}{Phys. Rev.
  A \textbf{68},  013820  (2003)}.

\bibitem{goldmanperiod}
N. Goldman and J. Dalibard, Periodically Driven Quantum Systems: Effective
  {Hamiltonians} and Engineered Gauge Fields,
  \href{https://doi.org/10.1103/PhysRevX.4.031027}{Phys. Rev. X \textbf{4},
  031027  (2014)}.

\bibitem{goldmandriven}
N. Goldman, J. Dalibard, M. Aidelsburger, and N.~R. Cooper, Periodically driven
  quantum matter: The case of resonant modulations,
  \href{https://link.aps.org/doi/10.1103/PhysRevA.91.033632}{Phys. Rev. A
  \textbf{91},  033632  (2015)}.

\bibitem{Bukovuniversal}
M. Bukov, L. D'Alessio, and A. Polkovnikov, Universal high-frequency behavior
  of periodically driven systems: from dynamical stabilization to Floquet
  engineering, \href{https://doi.org/10.1080/00018732.2015.1055918}{Adv. Phys.
  \textbf{64},  139  (2015)}.

\bibitem{Eckardthigh}
A. Eckardt and E. Anisimovas, High-frequency approximation for periodically
  driven quantum systems from a Floquet-space perspective,
  \href{https://doi.org/10.1088/1367-2630/17/9/093039}{New J. Phys.
  \textbf{17},  093039  (2015)}.

\bibitem{MANANGA}
E.~S. Mananga and T. Charpentier, On the {Floquet--Magnus} expansion:
  Applications in solid-state nuclear magnetic resonance and physics,
  \href{https://doi.org/10.1016/j.physrep.2015.10.005}{Phys. Rep. \textbf{609},
   1  (2016)}.

\bibitem{burgarth2024}
D. Burgarth, P. Facchi, A. Hahn, M. Johnsson, and K. Yuasa, Strong error bounds
  for {Trotter} and strang-splittings and their implications for quantum
  chemistry, \href{https://doi.org/10.1103/PhysRevResearch.6.043155}{Phys. Rev.
  Research \textbf{6},  043155  (2024)}.

\bibitem{YiKimMarvian2024}
C. Yi, L. Kim, and M. Marvian, Faster Randomized Dynamical Decoupling,
  \href{https://doi.org/10.48550/arXiv.2409.18369}{arXiv:2409.18369 [quant-ph]
  (2024)}.

\bibitem{BLANESmagnus}
S. Blanes, F. Casas, J.~A. Oteo, and J. Ros, The {Magnus} expansion and some of
  its applications, \href{https://doi.org/10.1016/j.physrep.2008.11.001}{Phys.
  Rep. \textbf{470},  151  (2009)}.

\bibitem{magnuson}
W. Magnus, On the exponential solution of differential equations for a linear
  operator, \href{https://doi.org/10.1002/cpa.3160070404}{Commun. Pure Appl.
  Math. \textbf{7},  649  (1954)}.

\bibitem{Blanespeda}
S. Blanes, F. Casas, J.~A. Oteo, and J. Ros, A pedagogical approach to the
  {Magnus} expansion, \href{https://doi.org/10.1088/0143-0807/31/4/020}{Eur. J.
  Phys. \textbf{31},  907  (2010)}.

\bibitem{Colombo2011}
F. Colombo, I. Sabadini, and D.~C. Struppa, Appendix: The Riesz--Dunford
  functional calculus,  in
  \href{https://doi.org/10.1007/978-3-0348-0110-2_5}{\textit{Noncommutative
  Functional Calculus: Theory and Applications of Slice Hyperholomorphic
  Functions}} (Springer Basel, Basel, 2011), pp.\ 201--210.

\end{thebibliography}
\end{document}